\documentclass[prd,nofootinbib,superscriptaddress]{revtex4}
\usepackage{amsmath}
\usepackage{breqn}
\usepackage{graphicx, subfig}
\usepackage{color}

\newcommand{\beq}{\begin{equation}}
\newcommand{\eeq}{\end{equation}}

\newcommand{\dd}{\partial}

\begin{document}

\title{Newton's Theorem of Revolving Orbits in Curved Spacetime}
\author{Pierre Christian}
\affiliation{Harvard-Smithsonian Center for Astrophysics$,$ 60 Garden Street$,$ Cambridge$,$ MA$,$ USA}
\date{02/01/2016}

\begin{abstract}
Newton's theorem of revolving orbits states that one can multiply the angular speed of a Keplerian orbit by a factor $k$ by applying a radial inverse cubed force proportional to $(1-k^2)$. In this paper we derive an extension of this theorem in general relativity, valid for the motion of massive particles in any static, spherically symmetric metrics. We verify the Newtonian limit of this extension and demonstrate that there is no such generalization for rotating metrics. Further we also extend the theory to the case of charged particles in the Einstein-Maxwell and Kaluza-Klein theories. 

\end{abstract}

\maketitle
\section{Introduction}
\begin{figure}
  \centerline{\includegraphics[scale=0.5]{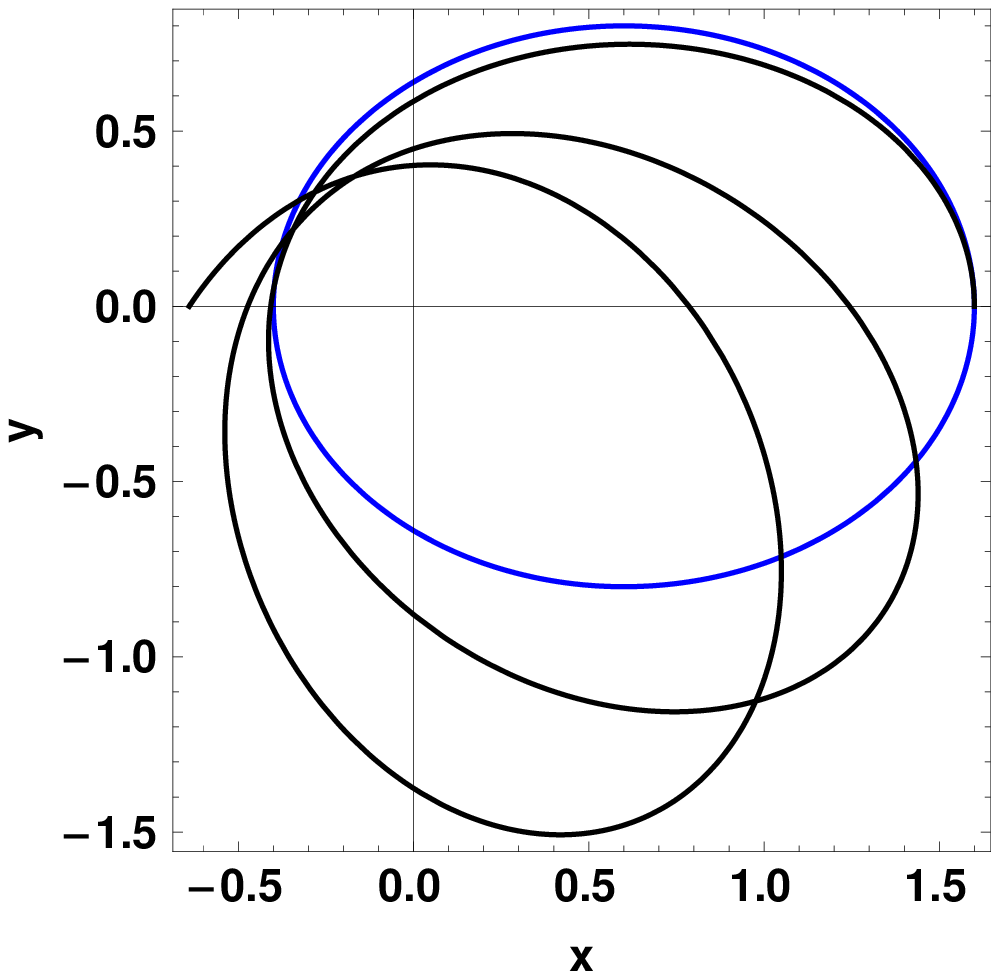}
   \includegraphics[scale=0.5]{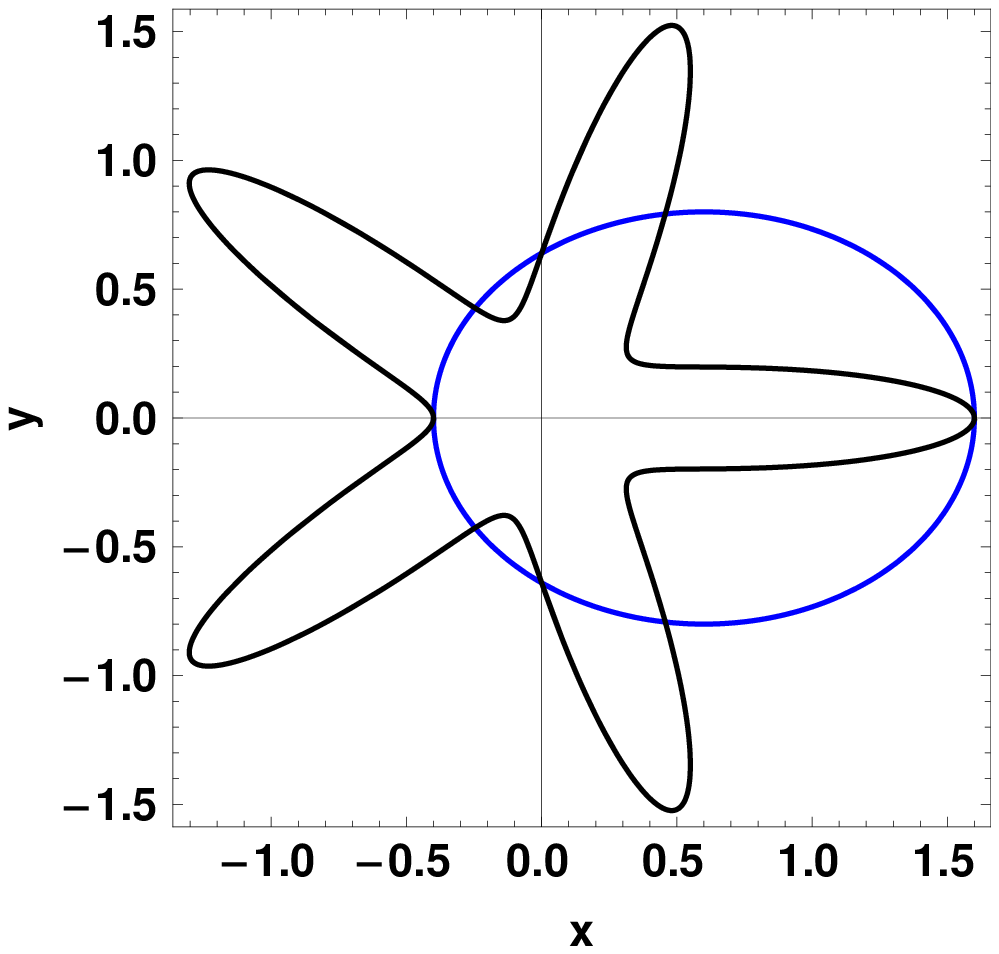} \includegraphics[scale=0.5]{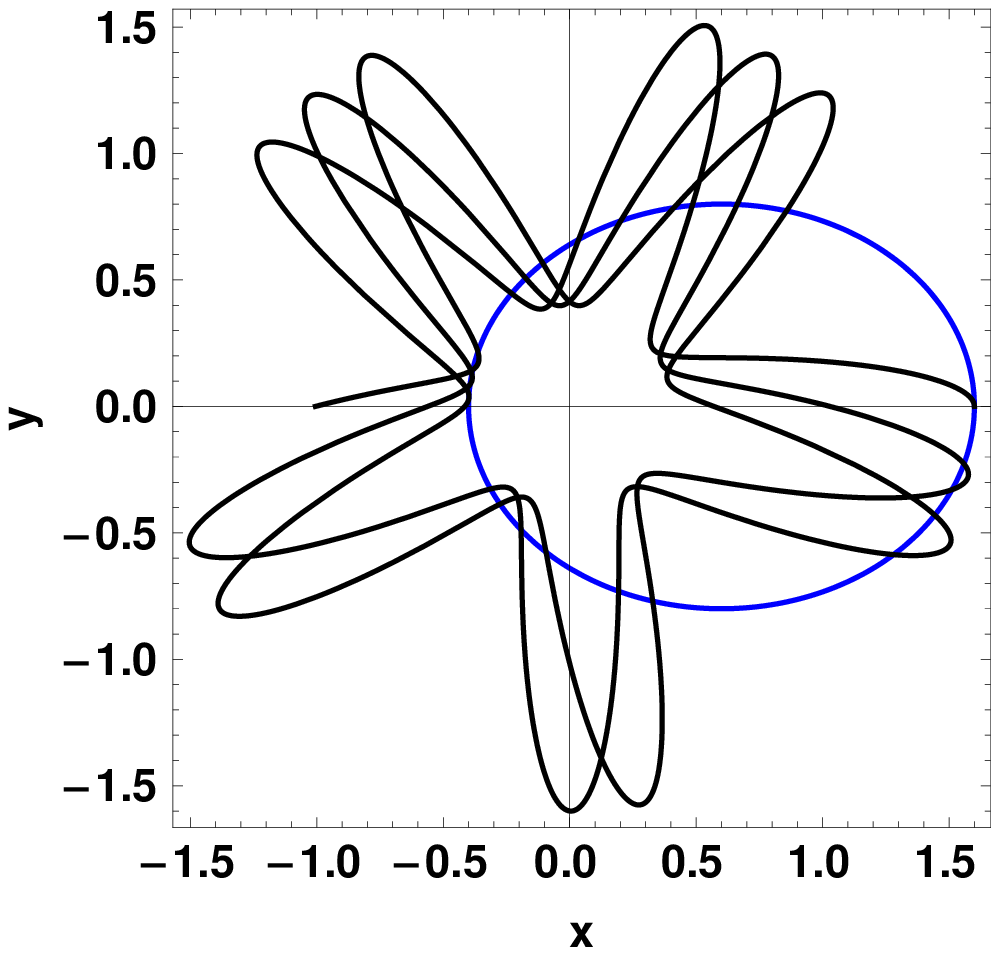}}
  \caption{\label{fig:newton_plots} Revolving orbits in Newtonian physics with a variety of $k$ factors (black lines). The blue ellipse is a standard,  Newtonian solution to the Kepler problem. When the absolute magnitude of $k$ is slightly off from unity (left figure), the ellipse fail to close upon itself, thus causing the orbit to undergo apsidal precession. This precession \emph{revolves} the orbit around the gravitating mass, located at $(0, 0)$. This revolution of the orbit can be clockwise ($k>1$) or counterclockwise ($k<1$). This behaviour is not unique for $k=1$. For $k$ a rational number, the orbit closed upon itself (middle figure). If $k$ is slightly different from rational, the orbit again fails to close upon itself, thus undergoing the aforementioned apsidal precession (right figure).} 
\end{figure}

In proposition 43-45 of the Principia \citep{Newton, Chandra}, Newton proved that the application of a radial force of the form 
\beq \label{eq:Newton}
F = \frac{L^2}{m r^3} (1-k^2) 
\eeq
to a particle of mass $m$ orbiting in a gravitational field with angular momentum $L$ will multiply its angular speed by a factor $k$ without changing its radial motion. These new orbits are called \emph{revolving} orbits because when $k$ is not a rational number, the new orbits will fail to close upon itself and an apsidal precession that revolves the orbit about the gravitating mass is induced. When $k$ is a ratio of integers, the orbits will close upon itself and produce exquisite patterns. However, when $k$ is changed slightly from this rational value, these patterns will also revolve about the gravitating mass. A  sample of revolving Keplerian orbits with a variety of $k$ values are plotted in Figure \ref{fig:newton_plots}.

Newton first develop this theorem in order to explain the apsidal precession of the moon. He used an extension of the theorem to prove that the moon's apsidal precession can be described either by the addition of a perturbing linear force (due to, ostensibly, the sun), or if gravitational force is modified so that its dependence to radius is an inverse power law with exponent $2 + 4/243$ instead of an inverse square law \citep{Chandra}. It is a historical curiosity that the first modification of Newtonian gravity is proposed by Newton himself, in the very book in which his law of gravitation is published.  

As noted by Chandrasekhar, the theorem of revolving orbits remains underdeveloped even $\sim$300 years after its publication. Donald Lynden-Bell \citep{DLB1,DLB2, DLB3, DLB4NZ} cited the theorem as a motivation for some of his work on classical dynamics, and the first extension came from Mahomed and Vawda \citep{MnV}, who described a generalization of the theorem where the radial motion between the old and the new orbits are not constrained to be the same.

Nguyen \citep{Nguyen} developed the first attempt to generalize this theorem to general relativity by deriving a revolving orbit theorem for the equation of motion of the Schwarzschild and de Sitter metrics. However, Nguyen did not use the full general relativistic equations and took inappropriate limits in deriving his results. We will return to this issue in the body of the paper.

In this paper we will develop a general relativistic extension to Newton's theorem of revolving orbits valid for massive particle motions in any static, spherically symmetric metric. In \S2 we provide modern proofs of the theorem of revolving orbits, in \S3 we derive the relativistic generalizations to the theorem, as well as verifying it in the Newtonian limit. In \S4 we shall demonstrate that there is no relativistic generalization of the theorem for rotating metrics, in \S5 we provide further extensions of the theorem for charged particles in electromagnetic fields and in Kaluza-Klein theory, and finally in \S6 we will provide some concluding remarks.

\section{Modern proofs of the theorem of revolving orbits}
Since the original proof by Newton relies on geometrical pictures that is unappealing to today's physicist, here we present two modern proofs of the theorem. The first, which we call the "force" derivation, is reproduced from Chandrasekhar's commentary of Newton's Principia \citep{Chandra}. Imagine a particle orbiting a Keplerian potential with angular speed $\omega$. If another particle orbits with the same radial motion, but with angular speed $k \omega$, where $k$ is some constant, then the angular momentum of the second particle is 
\beq \label{eq:kL}
L_2 = k L_1 \;,
\eeq
where $L_1$ is the angular momentum of the original particle. If the two particles' radial motion is the same then 
\beq
\frac{d^2 r}{dt^2} = F_1(r) + \frac{L_1^2}{m r^3} = F_2(r) + \frac{L_2^2}{m r^3} \; ,
\eeq
where $F_1(r)$ and $F_2(r)$ are the central forces (including gravity) applied on the first and second particles, respectively.  Rearranging the equation and using equation (\ref{eq:kL}), we obtain 
\beq
F_2 = F_1 + \frac{L_1^2}{m r^3} (1-k^2) \; .
\eeq
The extra $1/r^3$ force can be thought of as the extra force required to keep the radial motion of the two particles the same. One can understand this by thinking of particles in circular orbits; without the second term, a particle orbiting with angular momentum $kL_1$ will orbit at larger radius (if $k>1$) or smaller radius (if $k<1$). The second term compensates for this radial motion, allowing a particle with angular momentum $kL_1$ to orbit at the same radius as a particle with angular momentum $L_1$. Now we present an alternative proof for Newton's theorem of revolving orbit using the language of effective potentials. The orbital energy of a particle orbiting in a gravitational field with potential $V_g(r)$ is
\beq
E  = \frac{1}{2} m (\dot{r}^2 + r^2 \dot{\phi}^2) - V_g(r) \; ,
\eeq
where ($r$, $\phi$) are spherical coordinates and the overdots refers to derivatives with respect to time. This equation can rearranged to give
\beq
\frac{1}{2} m \dot{r}^2 = E - V_{\rm{eff}}(r) 
\eeq
where the effective potential is given by 
\beq \label{eq:NewtOriginalPotential}
V_{\rm{eff}}(r) = \frac{L^2}{2 m r^2} + V_g(r) \; .
\eeq
Applying equation (\ref{eq:kL}), we can see that the effective potential of the second particle can be written as
\beq \label{eq:NewtPotential}
V_{\rm{eff},2}(r) =V_{\rm{eff},1}(r)  + \frac{L_1^2}{2 m r^2} (k^2-1) \; .
\eeq
Taking the negative of the radial derivatives of $V_{\rm{eff},2}(r)$ gives the force applied to the second particle
\begin{align} \label{eq:ForceNewton}
F_2 &= -\frac{d V_{\rm{eff},2}(r) }{ d r} = \frac{L_1^2}{m r^3} - \frac{d V_g}{d r} - \frac{L_1^2}{m r^3} (k^2-1)
\\ &= F_1 + \frac{L_1^2}{m r^3} (1-k^2)\;,
\end{align}
again reproducing the theorem of revolving orbits. In the next section we will generalize these derivations relativistically. From this point onwards we take $c=G=1$, and our metric signature is $(-,+,+,+)$. 

\section{Relativistic generalization of the Revolving Orbit Theorem}
In this section we develop the relativistic generalization to the revolving orbit theorem by generalizing the force derivation to the general relativistic four-forces. Here we specialize to a spherically symmetric metric, given generally by the line element
\beq
ds^2 = - e^{2 \alpha(r)} dt^2 + e^{-2 \alpha(r)} dr^2 + r^2 d\Omega^2 \; ,
\eeq
where the metric function $\alpha(r)$ is a function of the radial coordinates only and $d\Omega^2$ is the 2-D round metric. This form of spherically symmetric metric is quite general, and is the form taken by a variety of famous metrics, including the Schwarzschild metric describing a non-spinning black hole with mass $M$, where
\beq
e^{2\alpha_{\rm{Sch}} (r)}= \left(1 - \frac{2 M}{r} \right)\; ,
\eeq
the Reissner-Nordstrom metric describing a non-spinning black hole with charge $Q$, where 
\beq \label{eq:Reiss}
e^{2\alpha_{\rm{RS}} (r)} = \left(1 - \frac{2 M}{r} +  \frac{Q^2}{r^2} \right) \;,
\eeq
and the de Sitter-Schwarzschild metric describing a non-spinning black hole embedded in a de Sitter universe with cosmological constant $\Lambda$, where
\beq \label{eq:Sitter}
e^{2\alpha_{\rm{dS-S}} (r)} = \left(1 - \frac{2 M}{r} - \frac{\Lambda}{3} r^2 \right) \; .
\eeq

In the Newtonian picture, "revolving" an orbit amounts to multiplying the angular speed with a factor $k$ while keeping the radial motion the same, i.e. 
\beq
\frac{\dd \phi}{\dd t} \rightarrow k \frac{\dd \phi}{\dd t} \; .
\eeq
In the relativistic picture, $\dot{\phi}$ and $\dot{r}$ are components of the velocity four vector $\mathbf{u}=[\dot{t}, \dot{r}, \dot{\theta}, \dot{\phi}]$, and revolving an orbit amounts to
\beq
[\dot{t}, \dot{r}, \dot{\theta}, \dot{\phi}] \rightarrow [\dot{t'}, \dot{r}, k\dot{\theta}, \dot{\phi}] \; ,
\eeq
where the overdots now correspond to derivative with respect to the proper time $\tau$. By the normalization condition $\mathbf{u} \cdot \mathbf{u} = -1$, the time component of the revolving velocity four vector is
\beq
\dot{t'} = \dot{t} \sqrt{1 + e^{-2\alpha} r^2 \Omega^2 (k^2-1)} \; ,
\eeq
where $\Omega$ is the coordinate angular velocity of the original, non-revolving orbit, $\Omega \equiv d \phi/d t$. For example, for the Schwarzschild metric, $\Omega^2=M/r^3$ for a circular orbit. The fact that the $\dot{t'} \neq \dot{t}$ reflects the notion that a particle moving with angular velocity $\dot{\phi}$ will have a different $dt/d\tau$ compared to a particle moving with angular velocity $k\dot{\phi}$.

Much like the case in flat spacetime, the general relativistic four-force (per unit mass), $f^\alpha$, causes a particle to deviate from its geodesic 
\beq \label{eq:4force}
\frac{d^2 x^\alpha}{d \tau^2} + \Gamma^{\alpha}_{\beta \gamma} \frac{d x^\beta}{d \tau} \frac{d x^\gamma}{d \tau} = f^\alpha \; .
\eeq
This is the relativistic generalization to the Newtonian equation $\vec{F}/m = \vec{a}$. Note that in the absence of four-force, equation (\ref{eq:4force}) reduces back to the usual geodesic equation, where a particle's four velocity is parallel transported along the geodesic. 

Evaluating the Christoffel symbols, the radial four-force equation of the spherically symmetric metric is given by
\beq \label{eq:rgeneral}
\frac{d^2 r}{d \tau^2} = \frac{\dd \alpha}{\dd r} \dot{t}^2 + \frac{\dd (-\alpha)}{\dd r} \dot{r}^2 - r e^{2 \alpha} \dot{\phi}^2 + f^r \;,
\eeq
where we have set $\theta=\pi/2$ with no loss of generality. Note that unlike the Newtonian derivation, now the force $f^r$ excludes gravity. Suppose there is a particle orbiting in such a spacetime with angular speed $\dot{\phi}$ without any external forces. This particle obeys the standard geodesic equation where $f^r=0$, 
\beq \label{eq:rmotionSpherical}
\frac{d^2 r}{d \tau^2} = \frac{\dd \alpha}{\dd r} \dot{t}^2 + \frac{\dd (-\alpha)}{\dd r} \dot{r}^2 - r e^{2 \alpha} \dot{\phi}^2 \; .
\eeq 
Suppose there is a second particle with four-velocity $\mathbf{u}'= [\dot{t'}, \dot{r}, k\dot{\theta}, \dot{\phi}] $ orbiting the spacetime under an external force $\mathbf{f}$.  The $r$-force equation for this particle reads
\begin{align}
\frac{d^2r}{d\tau^2} &= \frac{\dd \alpha}{\dd r} \dot{t'}^2 + \frac{\dd (-\alpha)}{\dd r} \dot{r}^2 - r e^{2 \alpha} k^2 \dot{\phi}^2 + f^r
\\ &=  \left[ 1 + e^{-2\alpha} r^2 \Omega^2 (k^2-1)  \right]  \frac{\dd \alpha}{\dd r} \dot{t}^2 + \frac{\dd (-\alpha)}{\dd r} \dot{r}^2 - r e^{2 \alpha} k^2 \dot{\phi}^2 + f^r \; .
\end{align}
Since the radial motion is identical, we can set  
\beq
\left[ 1 + e^{-2\alpha} r^2 \Omega^2 (k^2-1)  \right]  \frac{\dd \alpha}{\dd r} \dot{t}^2 + \frac{\dd (-\alpha)}{\dd r} \dot{r}^2 - r e^{2 \alpha} k^2 \dot{\phi}^2 + f^r = \frac{\dd \alpha}{\dd r} \dot{t}^2 + \frac{\dd (-\alpha)}{\dd r} \dot{r}^2 - r e^{2 \alpha} \dot{\phi}^2 \; .
\eeq
This gives the r-component of the four-force required to sustain the motion of the second particle,
\beq
f^r = r e^{2 \alpha} \dot{\phi}^2 (k^2-1) - e^{-2\alpha} r^2 \Omega^2 (k^2-1) \frac{\dd \alpha}{\dd r} \dot{t}^2 \; .
\eeq
We can rewrite this equation in terms of the angular momentum and energy of the first particle by noting that for static spherically symmetric spacetimes with a spacelike Killing vector $\boldsymbol\eta= (0,0,1,0)$, the angular momentum of a particle with four velocity $\mathbf{u}$ is given by
\beq
l = \boldsymbol\eta \cdot \mathbf{u} = r^2 \dot{\phi} \; ,
\eeq
where in Newtonian language $l$ is the angular momentum per unit mass, $l=L/m$. Similarly, the timelike Killing vector $\boldsymbol\xi = (1,0,0,0)$ gives the energy of the original particle to be 
\beq
e =-  \boldsymbol\xi \cdot \mathbf{u} = e^{2 \alpha} \dot{t} \; ,
\eeq
This means that a particle's angular speed can be multiplied by a factor $k$ without changing its radial motion if an external four-force whose r-component is 
\beq \label{eq:G1}
f^r = e^{2 \alpha} \frac{l^2}{r^3} (k^2-1) - e^{-6\alpha} r^2 e^2 \Omega^2 (k^2-1) \frac{\dd \alpha}{\dd r} \; ,
\eeq
is applied to it. This is the general relativistic version of Newton's theorem of revolving orbits. In the Newtonian version of the theorem, the motion of the second particle is obtained by adding a force to the first particle that depends on the angular momentum of the first particle $l$ and the angular speed multiplier $k$. In the relativistic picture, the motion of the second particle is produced by adding an extra four-force, which now depends not only on the momentum of the original particle $l$ and the multiplicative factor $k$, but also on the original particle's energy $e$. 

A four-force also obeys the constrain $\mathbf{f} \cdot \mathbf{u}=0$, which results in it having a time component,
\beq \label{eq:forceT}
f^t = \left[ \frac{e^{-2\alpha} l^2 (k^2-1)  }{r^3 } - e^{-10\alpha} r^2 e^2 \Omega^2 (k^2-1) \frac{\dd \alpha}{\dd r}  \right] \frac{\dot{r}}{\dot{t}} \; .
\eeq

Now, replacing $\dot{t} \rightarrow \dot{t'}$ in the $0$th component of the geodesic equation will give the time coordinate motion of a revolving orbit with $\mathbf{u'} =  [\dot{t'}, \dot{r}, k\dot{\theta}, \dot{\phi}] $,
\begin{align} \label{eq:TrueTimeMotion}
\frac{d^2 t'}{d\tau^2} &= - \Gamma^t_{rt} \dot{t'} \dot{r} 
\\ &= -\sqrt{1 + e^{-2\alpha} r^2 \Omega^2 (k^2-1)} \frac{\partial \alpha}{\partial r} \dot{t} \dot{r} \; .
\end{align}
Certainly this will result in a time coordinate motion that is different than the one prescribed by subtracting equation (\ref{eq:forceT}) to the original motion,
\begin{align} \label{eq:TrueRevolvingTimeMotion}
\frac{d^2 t'}{d\tau^2} &= \frac{d^2 t}{d\tau^2} - f^t
\\ &= - \frac{\partial \alpha}{\partial r} \dot{t} \dot{r} - f^t \; . 
\end{align}

Therefore, a four-force is incapable of sustaining a "true" revolving orbit, in which the time component motion $t(\tau)$ behaves as in equation (\ref{eq:TrueTimeMotion}). However, the orbital dynamics in the spatial dimensions, $r(\tau)$ and $\phi(\tau)$ are that of a revolving orbit. This means that adding the extra four-force revolves the orbit in the sense that the orbital shape, $r(\phi)$ as displayed in Figure \ref{fig:einstein_plots}, are correctly that of a revolving orbit, but the particle's position as a function of coordinate time, $r(t)$ and $\phi(t)$, is different between a particle under the influence of the extra four-force and a bona-fide revolving orbit due to their different $t(\tau)$ motions. 

Depending on the problem one is trying to solve, one might just be interested in revolving the spatial three dimensional orbit, $r(\phi)$, in which case there are no further issues. If one is also trying to determine the motion as a function of coordinate time, $r(t)$ and $\phi(t)$, accurately, a simple fix to this problem is by evolving the time component with equation (\ref{eq:TrueRevolvingTimeMotion}) instead of equation (\ref{eq:TrueTimeMotion}). However, note that such a fix is not truly covariant. One needs to first choose a frame before finding $t(\tau)$ in that frame using equation (\ref{eq:TrueRevolvingTimeMotion}), and the calculation has to be redone whenever one changes their frame. 

\subsection{Schwarzschild metric and the Newtonian limit}
Evaluating equation (\ref{eq:G1}) for the Schwarzschild metric gives
\begin{align}
f^r &= \left(1 - \frac{2 M }{r} \right) \frac{l^2}{r^3} (k^2-1) +  e^2 \Omega^2 (k^2-1) M \left(1 - \frac{2 M}{r} \right)^{-4}  
\\ &=  \left(1 - \frac{2 M }{r} \right) \frac{l^2}{r^3} (k^2-1)  + \frac{l^2}{r^3} (k^2-1) \frac{M}{r} + O\left[  \left( \frac{M}{r} \right)^3 \right] \; .
\end{align}
From equations (\ref{eq:Reiss}) and (\ref{eq:Sitter}), we can see that this is also the low order corrections to the revolving orbit theorem for both charged and de Sitter black holes under astrophysically relevant cases, where black holes possess very little charge $Q \ll M$ and the cosmological constant is small $\Lambda r^2 \ll 1$. 

\begin{figure}
  \centerline{\includegraphics[scale=0.7]{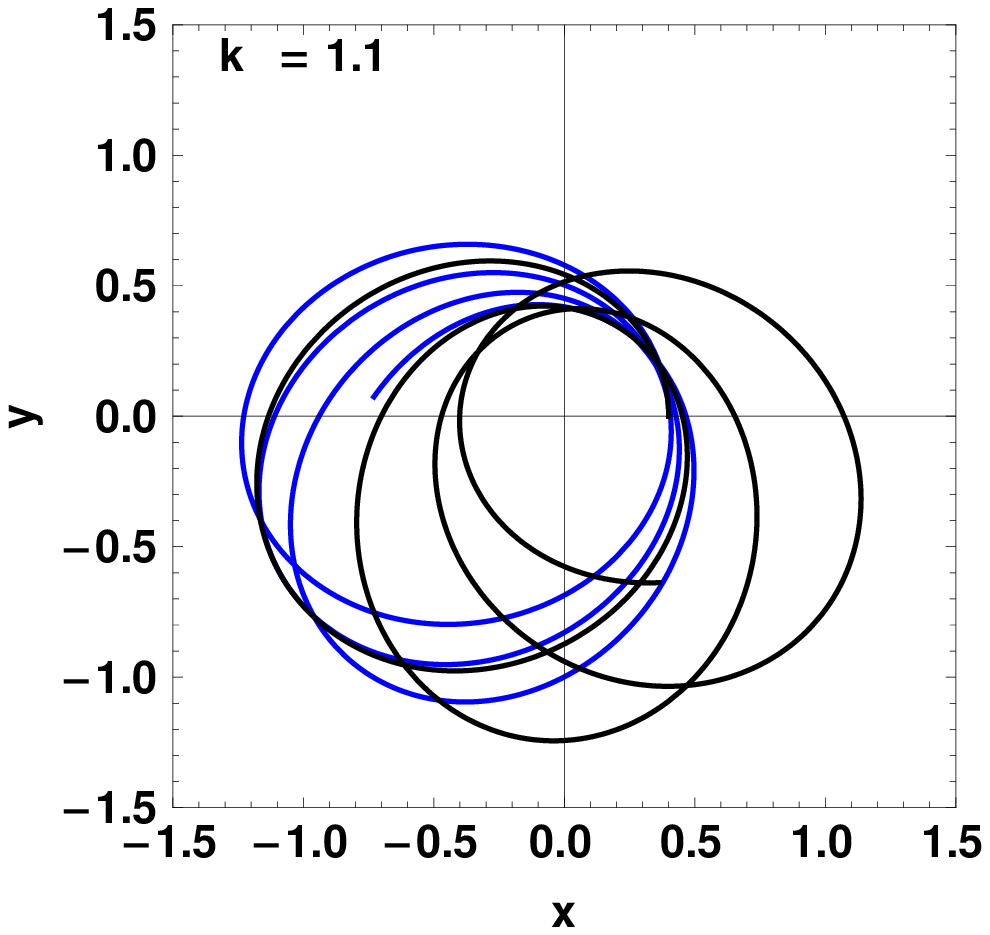}
  \includegraphics[scale=0.7]{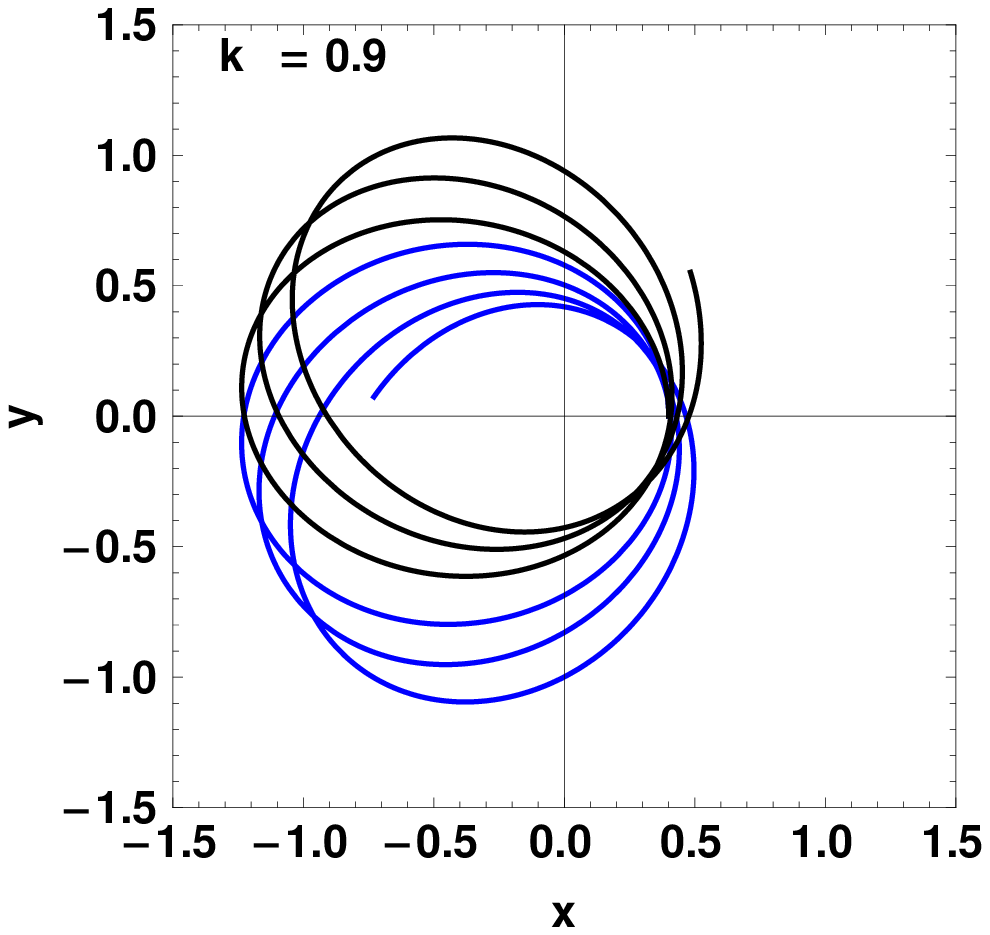}}
\centerline{\includegraphics[scale=0.7]{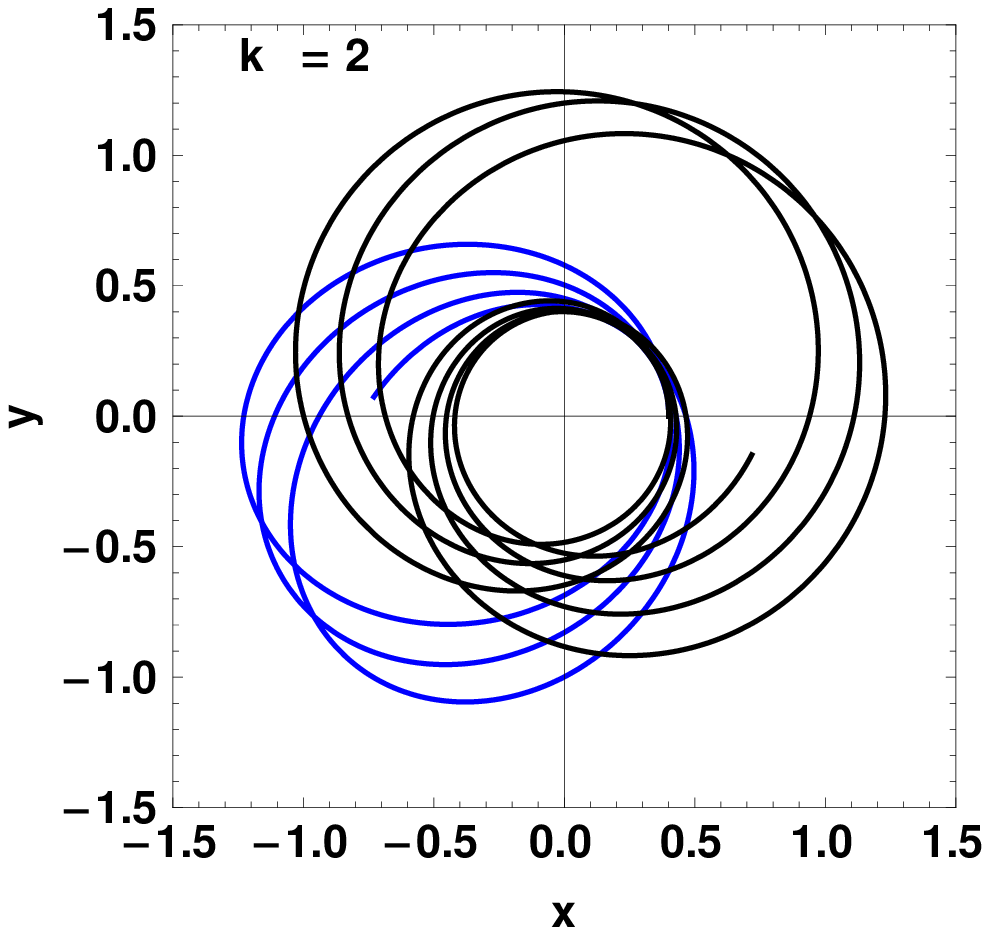}
\includegraphics[scale=0.7]{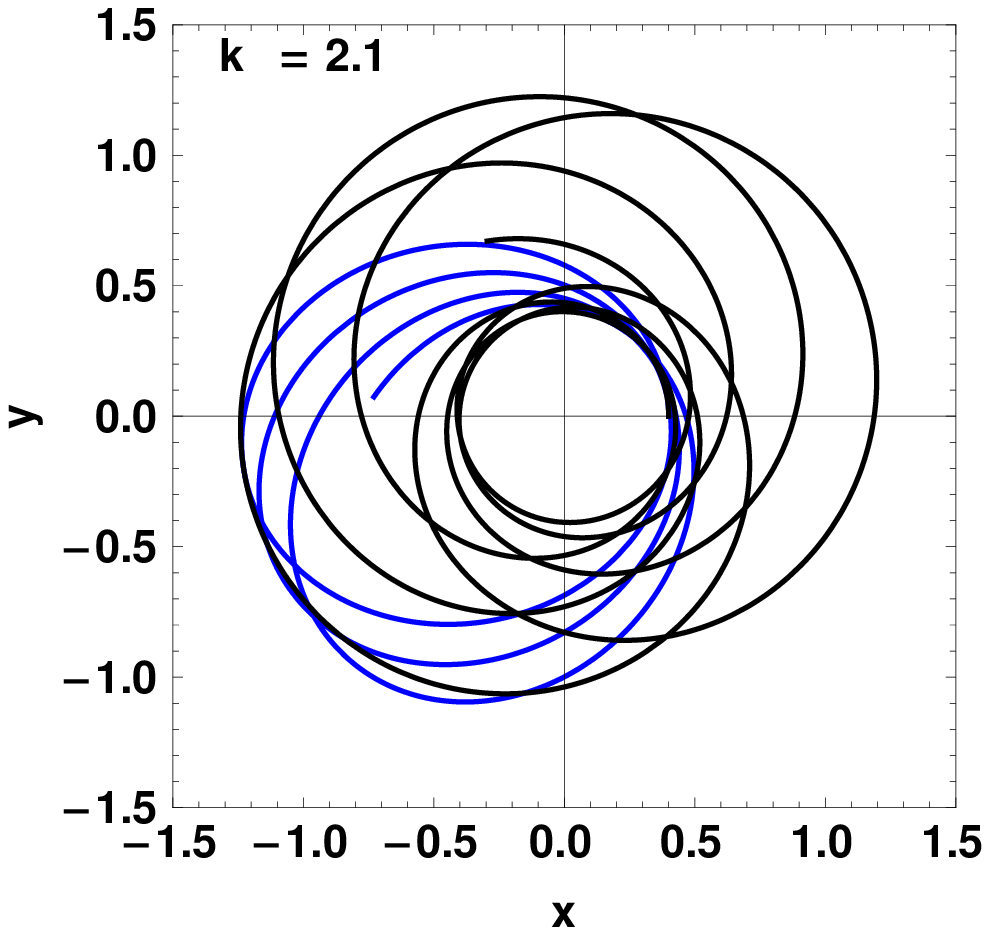}}
  \caption{\label{fig:einstein_plots} Revolving orbits in the Schwarzschild spacetime with a variety of $k$ factors (black lines). The blue line is a standard bound geodesic of a massive particle around a Schwarzschild black hole located at $(0, 0)$. The major difference between the Schwarzschild and the Newtonian case is that in the Schwarzschild metric the non-revolving solution is already undergoing apsidal precession. When $k$ is slightly off from unity, the orbit undergoes additional apsidal precession on top of the relativistic precession. As in the Newtonian case, this precession revolves the blue orbit around the gravitating mass in a clockwise fashion (if $k>1$) or in a counterclockwise fashion (if $k<1$). Due to the relativistic precession, rational values of $k$ do not close upon itself.} 
\end{figure}

Comparing this with the Newtonian version, equation (\ref{eq:Newton}), and noting that $l = L/m$, we found that the lowest order relativistic correction to the relativistic orbit theorem is just the extra addition of the term  
\beq \label{eq:RelCorrect}
f^r_{\rm{missing}} = - \left( \frac{2 M }{r}\right) \frac{L^2}{m r^3} (k^2-1) 
\eeq
to equation (\ref{eq:Newton}). It is clear why Newton missed this term: his equation of motion is valid when $r \gg M$, exactly the limit in which this relativistic correction disappears. A sample of revolving orbits with a variety of $k$ is plotted in Figure \ref{fig:einstein_plots}. The non-revolving Schwarzschild orbit (one with $k=1$) itself is already undergoing apsidal precession. When $k$ is slightly off from unity, an extra apsidal precession is added to the orbit. Much like the Newtonian case, this extra precession revolves the non-revolving orbit about the gravitating mass. 

However, one has to be careful when applying this correction. The Newtonian limit is obtained when motion of the orbiting particle is much smaller than the speed of light, meaning that $\dot{t} \rightarrow 1$. This limit is not independent to the limit $r \gg M$ because the closer a particle is to the black hole, the larger the particle's orbital velocity will be. This relation is made manifest by the equation
\beq \label{eq:tdot}
\dot{t} = \frac{e}{1 - \frac{2 M }{r}} \; ,
\eeq
where $e$ is the conserved quantity corresponding to the timelike Killing vector $\boldsymbol\xi = (1,0,0,0)$. To make its physical content in the Newtonian limit explicit, one can rewrite $e$ as 
\beq
e = \frac{E_N + m}{m} \; ,
\eeq
where $E_N$ is the Newtonian energy and the $m$ term corresponds to the rest mass energy of the particle. In the Newtonian limit, $E_N \sim v^2 \ll c^2$, thus $e \rightarrow 1$. Therefore, from (\ref{eq:tdot}) it is clear that $\dot{t} \rightarrow 1$ requires $r \gg M$.

Evaluating the Christoffel symbols of the Schwarzschild metric gives
\begin{equation} \label{eq:SchRadial}
\frac{d^2 r}{d \tau^2} = - \frac{M}{r^2} \left( 1- \frac{2M}{r} \right) \dot{t}^2 + \frac{M}{r^2} \left( 1- \frac{2M}{r} \right)^{-1} \dot{r}^2 + \left(1 - \frac{2 M }{r} \right) \frac{l^2}{r^3} \; ,
\end{equation}
and taking the limits $r \gg M$ and switching from proper to coordinate time we find to first order in $M/r$ 
\beq
\frac{d^2 r}{d t^2} = -\frac{M}{r^2} + \frac{M}{r^2} \frac{l^4}{2 r^4} + \left(1 - \frac{6 M}{r} \right) \frac{l^2}{r^3} \; .
\eeq
While the first term describes Newtonian gravity, and the $1$ in the parantheses is the Newtonian angular momentum barrier, this is not yet the radial Newtonian equation of motion. To fully reduce this to Newtonian, we have to impose another limit: that the angular momentum is small ($l^2 \ll r^2$), and in particular that $l^2/r^2$ is the same order as $M/r$. This is because the relativistic limit is also the slow motion limit, i.e. $\dot{t} \rightarrow 1$ implies $\dot{t}^2 \gg \dot{r}^2 + r^2 \dot{\phi}^2$, which implies $l^2/r^2 \ll 1$. At this limit we re-obtain the Newtonian equation
\beq
\frac{d^2 r}{d t^2} = -\frac{M}{r^2} + \frac{l^2}{r^3} \; .
\eeq
The point of this expansion is to show that one cannot simply append equation (\ref{eq:RelCorrect}) to equation (\ref{eq:Newton}), and then use Newtonian dynamics (first order in $M/r \sim l^2/r^2$) to evolve the particle's orbit. This relativistic correction enters at second order in $M/r$, and to be consistent, the evolution equation has to also be at least 2nd order in $M/r$ (i.e. equation (\ref{eq:SchRadial}) expanded the 2nd order $M/r$). This creates doubt to the results of \cite{Nguyen}, which uses the Newtonian equation of motion to claim that general relativistic precession can be cast in the form of Newton's revolving orbit theorem.

\section{Violation of the theorem for rotating metrics}
In this section we will show that there is no relativistic generalization of the theorem for rotating metrics, the most famous of which is the Kerr metric describing a black hole of mass $M$ and spin $a$,
\begin{equation}
ds^2 = -\left( 1- \frac{ 2 M r}{\rho^2} \right) dt^2 + \frac{\rho^2}{\Delta} dr^2 +\rho^2 d\theta^2+ \frac{\sin^2 \theta}{\rho^2} \left[ (r^2 + a^2)^2 - a^2 \Delta \sin^2 \theta \right] d \phi^2 - \frac{2 M r a \sin^2 \theta}{\rho^2} (dt d\phi + d\phi dt) \; , 
\end{equation}
where $\rho$ and $\Delta$ are defined as
\begin{align}
\rho^2 &\equiv r^2 + a^2 \cos^2 \theta \nonumber
\\ \Delta &\equiv r^2 - 2 M r + a^2  \; . \nonumber
\end{align}

In a spherically symmetric metric, conservation of angular momentum demands that orbits are planar. In the case of a rotating metric, however, angular momentum is only conserved along the symmetry axis. As such, in a rotating metric, the geodesics are not necessarily confined to a plane.

For example, in general rotating metrics carry gravitomagnetic fields that causes the Lense-Thirring precession. Particles orbiting with an orbital angular momentum vector $\mathbf{L}$ that is not parallel to the spin axis will have their $\mathbf{L}$ vector precess around the spin axis, causing the particle to undergo nodal precession. 

The first difficulty in generalizing the theorem of revolving orbits to rotating metrics is this additional degree of freedom. In the spherically symmetric and Newtonian case, one can multiply the angular speed by a factor $k$ while keeping the $r$ motion the same. For rotating metrics, multiplying angular speed by a factor $k$ can produce different effects depending on the orientation of the orbit with respect to the spin axis and the constraints on the $\theta$ motion.

In order to sidestep this difficulty we shall restrict our analysis to the equatorial plane, upon which the orbit is again planar. We will show that even in this highly symmetrical configuration the revolving orbit theorem will be violated for rotating metrics. The $\phi$ component of the geodesic equation for a rotating metric takes the form
\beq
\frac{1}{2} \frac{d^2 \phi}{d \tau^2} = - \Gamma^{\phi}_{r \phi} \dot{r} \dot{\phi} - \Gamma^{\phi}_{t r} \dot{r} \dot{t}  \; .
\eeq
In the case of a spherical spacetime, the second term in the right hand side is absent. Therefore, multiplying $\dot{\phi}$ and $\ddot{\phi}$ by $k$ gives an equation of motion for $\phi$ that is identical to that of a non-revolving orbit. This means that if at one point in time $\dot{\phi'}(\tau_0) = k \dot{\phi} (\tau_0)$, then at all times $\dot{\phi'}$ will be equal to $k \dot{\phi}$. This allows us to substitute $\dot{\phi'} = k \dot{\phi}$ in the other components of the geodesic equation. In general we cannot keep the revolving orbit, $\dot{\phi'}$, to be equal to $k$ times the non-revolving orbit, $\dot{\phi}$ at all times unless we add an extra force $f^{\phi}$ to the $\phi$ geodesic equation. However, Newton's theorem of revolving orbit demands that the 3-force must be radial. 

It could be said that the theorem of revolving orbits can be extended to cases where the 3-force is not confined to be radial, however such extension is beyond the scope of this work. Newton's original question: whether there is a radial (three) force that could turn a motion with angular speed $\dot{\phi}$ to one with angular speed $k \dot{\phi}$ without altering its radial motion has to be answered in the negative for the case of rotating metrics.

Another argument for the violation of the revolving orbit theorem for rotating metrics is to note that in the weak field limit general relativity reduces to a Lorentz force like equation \citep{Mashoon1}
\beq
\vec{F} = - \vec{E}_{g} - 2 \vec{v} \times \vec{B}_g \; ,
\eeq
where $\vec{E}_g$ is Newtonian gravity, $B_g$ is the gravitomagnetic field, and $\vec{v}$ the particle's velocity. As there is no theorem of revolving orbit for the Lorentz force in classical physics, there is no theorem of revolving orbit for the rotating metric in the weak field limit.

\section{Extension to electromagnetic and Kaluza-Klein theory}

\subsection{Theorem of revolving orbits in spacetimes with electromagnetic fields}
The theorem of revolving orbits can be extended to charged particles in spacetimes with electric fields if the electric fields obeys a spherical symmetry. The force equation in this case is given by
\beq
\frac{d^2 x^\alpha}{d \tau^2} + \Gamma^{\alpha}_{\beta \gamma} \frac{d x^\beta}{d \tau} \frac{d x^\gamma}{d \tau} = q F^{\alpha \gamma} g_{\gamma \kappa} \frac{d x^{\kappa}}{d \tau} \; ,
\eeq
where $q$ is the charge of the particle and the $F^{\alpha \beta}$ is the antisymmetric electromagnetic field strength tensor. In the case without magnetic fields, this equation in a spherically symmetric metric reduces to 
\beq
\frac{d^2 x^\alpha}{d \tau^2} + \Gamma^{\alpha}_{\beta \gamma} \frac{d x^\beta}{d \tau} \frac{d x^\gamma}{d \tau} = q F^{\mu 0} g_{00} \dot{t} \; .
\eeq
If the electric field is purely radial, the only equation that is modified is 
\beq
\frac{d^2 r }{d \tau^2} + \Gamma^{r}_{\beta \gamma} \frac{d x^\beta}{d \tau} \frac{d x^\gamma}{d \tau} = q F^{r0} g_{00} \dot{t} \; ,
\eeq
where $F^{r0}$ depends linearly on the radial electric field. As in the case without electromagnetic fields, we calculated the $r$ equation of motion for a revolving orbit, which now reads 
\begin{align}
\frac{d^2r}{d\tau^2} &= \frac{\dd \alpha}{\dd r} \dot{t'}^2 + \frac{\dd (-\alpha)}{\dd r} \dot{r}^2 - r e^{2 \alpha} k^2 \dot{\phi}^2 + q F^{r0} g_{00} \dot{t'} + f^r
\\ &=  \left[ 1 + e^{-2\alpha} r^2 \Omega^2 (k^2-1)  \right]  \frac{\dd \alpha}{\dd r} \dot{t}^2 + \frac{\dd (-\alpha)}{\dd r} \dot{r}^2 - r e^{2 \alpha} k^2 \dot{\phi}^2 -  \sqrt{ 1 + e^{-2\alpha} r^2 \Omega^2 (k^2-1) }  q F^{r0} e^{2\alpha} \dot{t} + f^r \; .
\end{align}
Comparing this with the equation for non-revolving orbit, and using the assumption that the $r(t)$ motion is the same, we obtain that now the force required to revolve the orbit is given by
\beq \label{eq:EMfr}
f^r = r e^{2 \alpha} \dot{\phi}^2 (k^2-1) - e^{-2\alpha} r^2 \Omega^2 (k^2-1) \frac{\dd \alpha}{\dd r} \dot{t}^2 -  \left[ 1 -  \sqrt{ 1 + e^{-2\alpha} r^2 \Omega^2 (k^2-1) }    \right] q F^{r0} e^{2 \alpha} \dot{t}  \; .
\eeq
If so desired, these equations can again be written in terms of the quantities $e$ and $l$, which are still constants of the motion.

In general, any configuration of magnetic fields will cause a violation of the theorem of revolving orbits. This is because If instead of a pure electric field, the particle is orbiting in a magnetic field, the radial force equation is given by 
\beq
\frac{d^2 r }{d \tau^2} + \Gamma^{r}_{\beta \gamma} \frac{d x^\beta}{d \tau} \frac{d x^\gamma}{d \tau} = q F^{r \phi} g_{\phi \phi} \dot{\phi} \; .
\eeq
Naively, we can perform the same exercise as before to obtain the force required to revolve the orbit. However, in the case of a magnetic field, the $\phi$ motion also changes, 
\beq
\frac{d^2 \phi}{ d \tau^2} = -2 \Gamma^{\phi}_{r \phi} \dot{r} \dot{\phi} + q F^{\phi r} \dot{r} \; .
\eeq
Revolving this orbit therefore does not lead to $\ddot{\phi} \rightarrow k \ddot{\phi}$, and instead requires the addition of a $4$-force whose $3$-force is not purely radial, thus violating the theorem of revolving orbits. If the component $F^{\phi r}$ is zero, the other components of the magnetic field will change the $\theta$ motion, since the magnetic force of the $\theta$ component reads
\beq
f^{\theta} = q F^{\theta r} g_{rr}\dot{r} + q F^{\theta \phi} g_{\phi \phi} \dot{\phi} \; .
\eeq
Once a $\theta$ motion is induced, the motion is no longer confined to a plane, and from the $\theta$ component of the force equation,
\beq
\frac{d^2 \theta}{d\tau^2} = -2 \Gamma^{\theta}_{r \theta} \dot{r} \dot{\theta} - 2 \Gamma^{\theta}_{\phi \phi} \dot{\phi}^2  + f^{\theta} \; ,
\eeq
it is clear that revolving this orbit requires an extra force in the $\theta$ direction, again violating the theorem of revolving orbits. 

\subsection{Theorem of revolving orbits in Kaluza-Klein theory}
The Kaluza-Klein theory \citep{Kaluza, Klein} is a five dimensional theory first proposed to unite gravity and electromagnetism in a geometric framework. It is a prototypical example of a field theory featuring a compact dimension, where one starts with a five dimensional manifold before compactifying one of the four space dimensions in a process often called dimensional reduction. The metric of the five dimensional manifold, $\tilde{g}_{AB}$, where capital letters $A$, $B$, etc ranges from $0$ to $4$ can be written in a line element as
\beq
\tilde{g}_{AB} dx^A dx^B= g_{\alpha \beta} dx^\alpha dx^\beta + \lambda^2 (A_\alpha dx^\alpha + dx^4)^2 \; , 
\eeq
where $A_\alpha$ is the electromagnetic $4$-potential, $\lambda$ a scalar field, and as before $g_{\alpha \beta}$ refers to the $4$ dimensional general relativistic metric. Note that $\alpha$, $\beta$ still ranges from $0$ to $3$. In this five dimensional manifold, a massive test particle obeys a $5$ dimensional geodesic equation,
\beq
\frac{d^2 x^A}{d \tilde{s}^2} + \Gamma^{A}_{BC} \frac{dx^A}{d\tilde{s}} \frac{dx^B}{d\tilde{s}} = 0 \; ,
\eeq
where $d\tilde{s}$ is the five dimensional proper time, $d\tilde{s} = - g_{AB} dx^A dx^B$. One can then proceed with dimensional reduction, assuming that the metric $g_{AB}$ does not depend on the coordinate $x^4$ and that the coordinate $x^4$ is compact and topologically a circle. This reduces the equation of motion into that of a four dimensional spacetime with some extra force terms \citep{KK1}
\beq
\frac{d^2 x^\alpha}{d \tau^2}  + \Gamma^{\alpha}_{\beta \gamma} \dot{x}^\beta \dot{x}^\gamma = q F^{\alpha \gamma} g_{\gamma \kappa} \dot{x}^\kappa + \frac{q^2 \lambda_\infty }{16 \pi \lambda^3 }\frac{d \lambda }{d x^\kappa} \left( g^{\alpha \kappa} + \dot{x}^\alpha \dot{x}^\kappa  \right) \; ,
\eeq
where $\lambda_\infty$ is a constant and now $\tau$ is the usual proper time in $4$ dimensional spacetime. Note that when $\lambda$ is a constant, the second term disappears and the Kaluza-Klein theory reduces to electromagnetism in curved spacetime.  

Let us consider a configuration where $F^{\alpha \beta} = 0$ and $\lambda$ is spherically symmetric, $\lambda = \lambda(r)$.  In this case it is possible to extend the theorem of revolving orbits to orbits in the Kaluza-Klein theory with a nontrivial $\lambda$. In such a configuration, the motion of the particle is again constrained within a plane. As before we can choose the $\theta=\pi/2$ plane without loss of generalization. The $\phi$ component of the motion is given by
\begin{align}
\frac{d^2 \phi}{d \tau^2} &= -2 \Gamma^{\phi}_{r \phi} \dot{r} \dot{\phi} + \frac{K}{\lambda^3} \frac{d \lambda}{d r} \dot{r}\dot{\phi} \nonumber
\\ &= \left[-2 \Gamma^{\phi}_{r \phi} \dot{r} + \frac{K}{\lambda^3} \frac{d \lambda}{d r} \dot{r} \right] \dot{\phi} \; ,
\end{align}
where $K \equiv q^2 \lambda_\infty/16 \pi$. This equation of motion is unchanged by replacing $\dot{\phi}$ with $\ddot{\phi}$. Therefore, unlike in the case of a rotating metric, if $\dot{\phi'} = k\dot{\phi}$ at one time $\tau_0$, the relation will be true for all times, as is necessary for the validity of the theorem of revolving orbits.  The $r$ component of the motion for a spherically symmetric metric is given by
\beq
\frac{d^2 r}{d \tau^2} =  \frac{\dd \alpha}{\dd r} \dot{t}^2 + \frac{\dd (-\alpha)}{\dd r} \dot{r}^2 - r e^{2 \alpha} \dot{\phi}^2 + \frac{K}{\lambda^3} \frac{d \lambda}{d r}(e^{2 \alpha} + \dot{r}^2) \;,
\eeq
which means if we revolve the orbit, $\mathbf{u} \rightarrow \mathbf{u'}$, we can calculate $f^r$ by 
\begin{align}
\frac{d^2r}{d\tau^2} &= \frac{\dd \alpha}{\dd r} \dot{t'}^2 + \frac{\dd (-\alpha)}{\dd r} \dot{r}^2 - r e^{2 \alpha} k^2 \dot{\phi}^2 + \frac{K}{\lambda^3} \frac{d \lambda}{d r}(e^{2 \alpha} + \dot{r}^2) + f^r
\\ &=  \left[ 1 + e^{-2\alpha} r^2 \Omega^2 (k^2-1)  \right]  \frac{\dd \alpha}{\dd r} \dot{t}^2 + \frac{\dd (-\alpha)}{\dd r} \dot{r}^2 - r e^{2 \alpha} k^2 \dot{\phi}^2 + \frac{K}{\lambda^3} \frac{d \lambda}{d r}(e^{2 \alpha} + \dot{r}^2) + f^r \; .
\end{align}
Because the extra term from Kaluza Klein is only a function of $r$, it is unchanged when we revolve the orbit by setting $\mathbf{u} \rightarrow \mathbf{u'}$. Therefore, to keep $r(\tau)$ identical with that of a non-revolving orbit, $f^r$ is exactly the same as the $f^r$ in the case of pure general relativity,
\beq
f^r = e^{2 \alpha} \frac{l^2}{r^3} (k^2-1) - e^{-6\alpha} r^2 e^2 \Omega^2 (k^2-1) \frac{\dd \alpha}{\dd r} \; .
\eeq
The theorem of revolving orbits works without modification if one adds a Kaluza-Klein field $\lambda(r)$. If the electromagnetic tensor $F^{\alpha \beta}$ is not zero and $\lambda=\lambda(r)$, then the force $f^r$ will instead be the same as that of the previous section's equation (\ref{eq:EMfr}).

\section{Conclusion}
In this paper we developed a relativistic generalization of Newton's theorem of revolving orbits valid for the motion of massive particles around any static, spherically symmetric metrics. We showed how this extension reduces back to the Newtonian formula at the appropriate limit, and how the theorem is violated in the case of a rotating metric. Finally we extend the theorem of revolving orbits to the case of charged particles in the Einstein-Maxwell and Kaluza-Klein theories. 

Much like the original Newtonian theorem of revolving orbits, these extensions are only valid in the case of orbiting test particles. When the orbiting particles are allowed to modify the metric, effects such as gravitational radiation reaction is introduced, and could result in the violation of the theorem even in the static, spherically symmetric case. 

\section{Acknowledgement}
The author would like to thank Avi Loeb for comments on the manuscript, and Ramesh Narayan for pointing out an error in the original manuscript.

\end{document}